\documentclass[twocolumn,showpacs,preprintnumbers,amsmath,amssymb]{revtex4}

\usepackage{graphicx}
\usepackage{dcolumn}
\usepackage{bm}

\begin{document}

\preprint{APS/123-QED}

\title{Energy of vanishing flow in heavy-ion collisions: Role of Coulomb interactions and asymmetry of a 
reaction \\} 

\author{Sanjeev Kumar, Varinderjit Kaur}
\author{Suneel Kumar}%
 \email{suneel.kumar@thapar.edu}
\affiliation{
School of Physics and Materials Science, Thapar University Patiala-147004, Punjab (India)\\}
\date{\today}
\begin{abstract}
We aim to understand the role of Coulomb interactions as well as of different equations of state 
on the disappearance of transverse flow for various asymmetric reactions leading to same total mass. 
For the present study, the total mass of the system is kept constant (${A_{TOT}}$ = 152) and asymmetry 
of the reaction is varied between 0.2 and 0.7. We find that the 
contribution of mean-field at low incident energies is more for nearly symmetric systems, 
while the trend is opposite at higher incident energies. The Coulomb interactions as well as different
equations of state are found to affect the balance energy significantly for larger asymmetric reactions.  
\end{abstract}
\maketitle
\section{Introduction}
The heavy-ion physics has attracted much attention during the last three decades \cite{scheid,aichelin,
carlen,chen,li}. The behavior of nuclear matter under the extreme conditions of temperature, density, 
angular momentum etc., is a very important aspect of heavy-ion physics. One of the important quantity 
which has been used extensively to study this hot and dense nuclear matter is the collective 
transverse in-plane flow \cite{scheid,aichelin,sood}.
This quantity has a beauty of vanishing at a certain incident energy. This energy is dubbed as balance 
energy (${E_{bal}}$) or the energy of vanishing flow (EVF) \cite{sood,krofcheck}. This beauty is due to 
the counterbalancing of attractive mean-field at low 
incident energies and repulsive nucleon-nucleon (NN) collisions at higher incident energies. The balance 
energy of the masses ranging from ${C^{12}+C^{12}}$ to ${U^{238}+U^{238}}$ at different colliding 
geometries was studied experimentally and theoretically and found to be sensitive with the
composite mass of the system \cite{sood,mota} as well as with the impact parameter of a reaction 
\cite{sood,lukasik,sullivan}.\\  
With the passage of time, isospin degree of freedom in terms of symmetry
energy and NN cross section is found to affect the 
balance energy or energy of vanishing flow and related phenomenon in heavy-ion collisions
\cite{chen,li,rpak,gautam}. \\
Experimentally, Pak {\it et al.}, studied the isospin effects on the collective flow and balance energy at 
central and peripheral collision geometries \cite{rpak}. On the other hand, theoretically, 
this effect is studied 
by using the isospin-dependent Boltzmann Uehling-Uhlenbeck model (IBUU) \cite{carlen,li,li2}, 
and isospin-dependent quantum molecular dynamics (IQMD) model \cite{lukasik,gautam, liewen,hartnack}.\\
As noted, balance energy is due to the counterbalancing of the attractive mean-field and repulsive nucleon-
nucleon collisions. The Coulomb interaction in intermediate energy heavy-ion collisions is expected
to play a dominant role in balance energy due to its repulsive nature. These effects are
supposed to be more pronounced in the presence of isospin effects \cite{liu}. The comparative 
study which will show the shift in balance energy due to Coulomb interactions in the presence of 
isospin effects by taking into account asymmetry of reaction in a controlled fashion is still missing 
in the literature. The second point is the asymmetry of the reaction. In some of the studies, the asymmetry
of a reaction is taken into care, but not in other, which is very important to study the isospin 
effects \cite{liu,zhang}. The asymmetry of the reaction can be
defined by the parameter ${{\eta}=  {\mid(A_T-A_P)}/{(A_T+A_P)\mid}}$; where ${A_T}$ and ${A_P}$ are the masses of
target and projectile. The ${\eta}$ = 0 corresponds to the symmetric reactions, whereas, non-zero
value of ${\eta}$ define different asymmetry of the reaction. It is worth mentioning that the 
reaction dynamics in a symmetric reaction (${\eta}$ = 0) can be quite different compared to 
asymmetric reaction (${{\eta} \ne 0}$) \cite{donangelo}. This is due to the deposition of excitation 
energy in the form
of compressional energy and thermal energy in symmetric and asymmetric reactions, respectively. The
effect of the asymmetry of a reaction on the multifragmentation is studied many times in the literature
\cite{liu,zhang,donangelo}. 
Unfortunately, very little study is available for the asymmetry of the reaction in terms of 
transverse in-plane flow.\\
In this paper, we will perform the first ever study for the balance energy in terms of asymmetry of the 
reaction and then observe the effect of Coulomb interactions, symmetry energy, equations of state as
well as different frame of references. The IQMD model used for the present analysis is explained in 
the Sec.-II. The results are presented in Sec.-III, leading to the conclusions in Sec.-IV. \\
\section{The Model}
The isospin-dependent quantum molecular dynamics (IQMD)\cite{hartnack} model treats different 
charge states of nucleons, deltas and pions explicitly \cite{hartnack2}, as inherited from the 
Vlasov-Uehling-Uhlenbeck (VUU) model 
\cite{kruse}. The IQMD model was used successfully in analyzing the large number of observables 
from low to relativistic energies \cite{gautam,hartnack,hartnack2,kruse}. One of its version (QMD), has been very successful in explaining 
the subthreshold particle 
production \cite{huang}, multi-fragmentation \cite{fuchs,PRC58}, collective flow \cite{sood,sood1}, 
disappearance of flow \cite{sood}, and density temperature reached in a reaction \cite{fuchs}. 
We shall not take relativistic effects into account, since in
the energy domain we are interested, there is no relativistic effect \cite{lehmann}. The isospin degree 
of freedom enters into the calculations via both cross sections, mean field and Coulomb interactions  
\cite{kruse}. The details about the elastic and inelastic cross sections for 
proton-proton and neutron-neutron collisions can be found in Refs.\cite{hartnack,lehmann}. \\
In this model, baryons are represented by Gaussian-shaped density distributions \\
\begin{equation}
f_i(r,p,t) = \frac{1}{{\pi}^2{\hbar}^2}e^{\frac{{-(r-r_i(t))^2}}{2L}}e^{\frac{{-(p-p_i(t))^2}.2L}{\hbar^2}}.
\end{equation}
Nucleons are initialized in a sphere with radius R = ${1.12A^{1/3}}$ fm, in accordance with the  liquid 
drop model. Each nucleon occupies a volume of ${\hbar^3}$ so that phase space is uniformly filled.
The initial momenta are randomly chosen between 0 and Fermi momentum ${p_F}$. The nucleons of the target 
and projectile interact via two and three-body Skyrme forces and Yukawa potential. The isospin degrees of 
freedom is treated explicitly by employing a symmetry potential and explicit Coulomb forces between 
protons of the colliding target and projectile. This helps in achieving the correct
distribution of protons and neutrons within the nucleus.\\
The hadrons propagate using Hamilton equations of motion:\\
\begin{equation}
\frac{d\vec{r_i}}{dt} = \frac{d<H>}{d{p_i}}~~~~;~~~~\frac{d\vec{p_i}}{dt} = -\frac{d<H>}{d{r_i}}.
\end{equation}
 with\\ 
${ <H> = <T> + <V>}$  is the Hamiltonian.
\begin{eqnarray}
   & =&  \sum_i\frac{p_i^2}{2m_i} + \sum_i \sum_{j>i}\int f_i(\vec{r},\vec{p},t)V^{ij}(\vec{r'},\vec{r})\nonumber\\ 
& &\times f_j(\vec{r'},\vec{p'},t)d\vec{r}d\vec{r'}d\vec{p}d\vec{p'}.
\end{eqnarray}
The baryon-baryon potential ${V^{ij}}$, in the above relation, reads as\\
\begin{eqnarray}
V^{ij}(\vec{r'}-\vec{r}) &~=~& V_{Skyrme}^{ij} + V_{Yukawa}^{ij} + V_{Coul}^{ij} + V_{Sym}^{ij}\nonumber\\
&=&t_1\delta(\vec{r'}-\vec{r})+ t_2\delta(\vec{r'}-\vec{r}){\rho}^{\gamma-1}(\frac{\vec{r'}+\vec{r}}{2})\nonumber\\ 
&+& t_3\frac{exp({\mid{\vec{r'}-\vec{r}}\mid}/{\mu})}{({\mid{\vec{r'}-\vec{r}}\mid}/{\mu})} + \frac{Z_{i}Z_{j}{e^2}}{\mid{\vec{r'}-\vec{r}\mid}}\nonumber\\
&+& t_{4} \frac{1}{\rho_o}T_{3}^{i}T_{3}^{j}.\delta(\vec{r'_i} - \vec{r_j}).
\end{eqnarray}
Where ${{\mu} = 0.4 fm}$, ${t_3 = -6.66 MeV}$ and ${t_4 = 100 MeV}$. The values of ${t_1}$ and ${t_2}$ 
depends on the values of ${\alpha}$, ${\beta}$ and ${\gamma}$ \cite{aichelin}.
Here ${Z_i}$ and ${Z_j}$ denote the charges of the ${i^{th}}$ and ${j^{th}}$ baryon, and ${T_{3}^i}$, 
${T_{3}^j}$ are their respective ${T_3}$ components (i.e. 1/2 for protons and -1/2 for neutrons). 
The parameters ${\mu}$ and ${t_1,........,t_4}$ 
are adjusted to the real part of
the nucleonic optical potential. For the density dependence of the nucleon optical potential, standard 
Skyrme-type parameterizations is employed. The Skyrme energy density have been shown to be very
successful at low incident energies where fusion is dominant channel \cite{rkpuri,gupta}. 
The Yukawa term is quite
similar to the surface energy coefficient used in the calculations of nuclear potential for fusion
\cite{ishwar}. 
The choice of the equation of state (or compressibility) is still a 
controversial one. Many studies advocates softer matter, whereas, much more believe the matter to be 
harder in nature \cite{sood,kruse}. We shall use both hard (H) and soft (S) equations of
state that have compressibilities of 380 and 200 MeV, respectively. \\
\section{Results and Discussions}
As discussed earlier, asymmetry of a reaction is found to affect the phenomena of muti-fragmentation in 
intermediate energy heavy-ion collisions \cite{zhang,donangelo}. On the other hand, system mass 
dependence of balance energy is studied many times in the literature \cite{sood}. To check the effect 
of Coulomb interactions and asymmetry of a
reaction on the balance energy, we have fixed (${A_{TOT}}$ = ${A_T +A_P}$ = 152) and varied the asymmetry 
of the reaction just like this: $_{26}Fe^{56}+_{44}Ru^{96}$ (${\eta = 0.2}$), $_{24}Cr^{50}+_{44}Ru^{102}$
(${\eta = 0.3}$), $_{20}Ca^{40}+_{50}Sn^{112}$ (${\eta = 0.4}$), $_{16}S^{32}+_{50}Sn^{120}$ 
(${\eta = 0.5}$), $_{14}Si^{28}+_{54}Xe^{124}$ (${\eta = 0.6}$), $_{8}O^{16}+_{54}Xe^{136}$ 
(${\eta = 0.7}$). The asymmetry of a reaction with multi-fragmentation in this fashion is varied 
many times \cite{zhang,donangelo}. The whole reaction dynamics is studied at semi-central geometry 
by varying the incident energy between 50 and 250 MeV/nucleon with an increment of 50 MeV/nucleon 
by employing hard as well as soft equations of state. We have checked the stability of the reacting 
nuclei in laboratory (lab) as well as in center of mass (c.m.) frame by taking into account the Coulomb 
interactions. Our main purpose here is to understand the effect of equations of state and Coulomb 
interactions on the energy of vanishing flow or alternatively, on the balance energy by taking into 
account the asymmetry of a reaction.\\
The directed transverse flow is calculated using ${<P_{x}^{dir}>}$ \cite{sood}\\
\begin{equation}
<P_{x}^{dir}> = \frac{1}{A}\sum_i {sgn}\{Y(i)\} P_{x}(i),
\end{equation}
where Y(i) and ${P_x(i)}$ are, respectively, the rapidity distribution and transverse momentum of the 
${i^{th}}$ particle.\\ 
\begin{figure}
\includegraphics{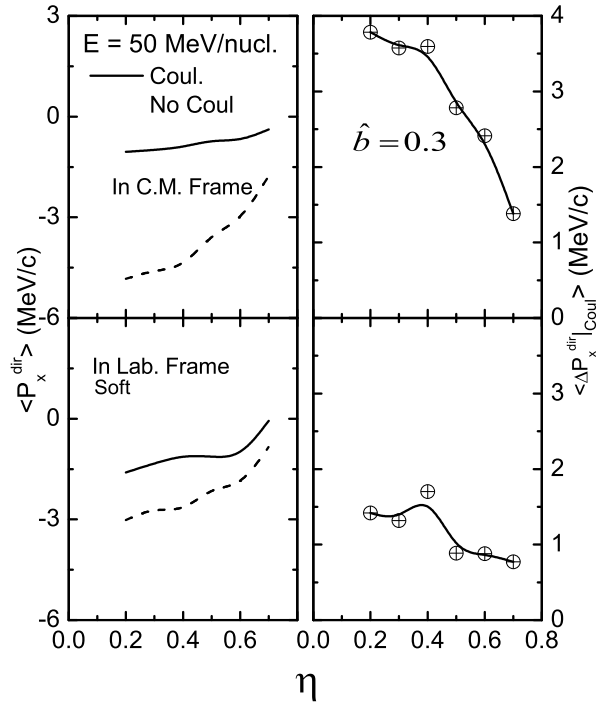}
\caption{\label{fig1}Asymmetry dependence of directed flow in lab as well as center of mass frame in L.H.S, while, 
R.H.S for the relative effect of Coulomb interactions. The different lines in the figure are 
representing the effect of symmetry energy and Coulomb interactions.}
\end{figure} 
To check the effect of frame of reference, we display in Fig.1, variation of the asymmetry ${\eta}$
on directed flow $<{P_{x}^{dir}}>$ in lab. as well as center of mass frame at incident energy of 
E = 50 MeV/nucleon.
The top and bottom panels in the right hand side of the figure are representing 
the relative Coulomb effect with respect to the asymmetry of the reaction. This relative effect is
calculated as:\\
\begin{equation}
<\Delta P_{x}^{dir}\mid_{Coul}> = <P_{x}^{dir}>_{Coul+Sym} - <P_{x}^{dir}>_{No Coul+Sym}
\end{equation}
As is evident from the figure, directed flow is found to increase in a systematic manner in C.M. as well 
as in  lab frame at E = 50 MeV/nucleon with asymmetry of the reaction. The inclusion of Coulomb 
interactions does not alter the conclusions. Note that the increase in the asymmetry is related with 
the increase in the N/Z ratio. Because the symmetry potential for
the neutron rich systems is stronger compared to the neutron poor systems due to large relative neutron
strength. Furthermore, the symmetry potential is repulsive for neutrons and attractive for protons. 
On the other hand, more negative value of directed flow (dominating the mean field) is observed in 
the absence of Coulomb interactions in center of mass as well as in lab. frames. This is due to the 
enhancement of the chemical 
and mechanical instability domains in the absence of Coulomb interactions \cite{coulomb}. Similar 
type of study and conclusion was also performed for nuclear stopping in ref. \cite{liu}. \\ 
Extensive study in the literature proves the stability of reactions in lab. frame, but, keep
in mind that the reaction under consideration in these studies were symmetric in nature. As is clear 
from the figure, asymmetric systems are found to be more stable in the center of mass frame compared 
to the lab frame. Moreover, if one consider lab frame, one is surely missing the effect of 
asymmetry. To further strengthen the stability of center of mass frame with asymmetry, the relative 
effect of Coulomb interactions ${<\Delta P_{x}^{dir}\mid_{Coul}>}$
is studied in both frames. The relative effect of the Coulomb interactions is found to decrease with 
increase in the asymmetry of a reaction. The 
systematic decrease can be seen in center of mass frame with asymmetry, while very weak dependence
of Coulomb interactions on the asymmetry is obtained in lab frame. For the further study, we have
opted the center of mass frame because we want to see the shift in the balance energy due to Coulomb 
interactions, whose effect is clearly visible in center of mass frame. \\
\begin{figure}
\includegraphics{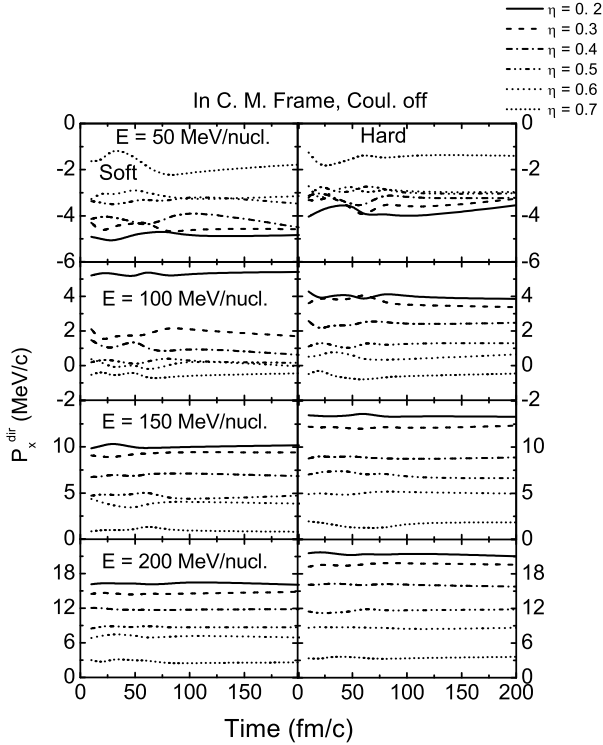}
\caption{\label{fig2} The time evolution of directed flow at different incident energies in center of mass frame in
the absence of Coulomb interactions. The left and right panels are for soft and hard equations of
state, respectively.}
\end{figure}
Before we proceed further, let us check the time evolution of directed transverse flow. In Fig. \ref{fig2}, we 
display the time evolution of directed flow from E = 50 to 200 MeV/nucleon in center of mass frame for 
Soft (L.H.S) and Hard (R.H.S) equations of state. Note that the compressibilities of soft and hard 
equations of state are 200 and 380 MeV, respectively \cite{aichelin}. The time evolution is plotted in 
the absence of Coulomb interactions to see the maximum effect of asymmetry of a reaction on the directed 
in-plane flow. The figure reveals the following points:\\
a). The quantity ${<P_{x}^{dir}>}$ is observed to be constant throughout the whole distribution of time, 
while, the large variation
is observed in the value at initial and final time steps when observed in the lab frame \cite{sood}\\
b). With the increase in the incident energy, the directed flow is approaching towards more positive value.
This is due to the well known fact of increase in the frequent NN collisions with increase in the 
incident energy \cite{sood}. \\
c). The behavior of the directed flow with asymmetry of a reaction follows the opposite trend at
E = 50 MeV/nucleon as compared to other high incident energies. It has been discussed many times in 
literature and also clear from the present findings that attractive mean-field is dominating at E = 50 MeV
/nucleon compared to higher incident energies under consideration \cite{sood,sanjeevCPL}. This is due 
to the different mechanisms contributing at low and high incident energies within isospin-dependent quantum molecular 
dynamics. It was shown by us as well as by others \cite{chen,li,gautam} that symmetry potential dominates 
the physics
at low incident energy, while, NN cross sections one major driven force at higher incident energies. Furthermore, at low incident 
energies, symmetry potential is repulsive for neutrons and attractive for
protons. With the increase in the asymmetry of a reaction, the number of neutrons increases and hence comparative 
repulsion due to neutrons also increases leading to less attractive value of the flow. It is also 
clear from ref. \cite{zhang}, that with the
increase in asymmetry of a reaction, the participant zone decreases and
spectator zone increases. At higher incident energies, where the effect of symmetry energy is negligible, the contribution of NN collisions comes from the participant zone whereas mean-field contribution comes
from the spectator zone.
Hence directed flow is found to be less positive with increase in the asymmetry of the reaction.\\
d). The directed flow has less positive value with soft equation of state compared to hard 
equation of state. The less positive means the dominance of mean-field. This is due to the different 
compressibilities of hard (380 MeV) and soft (200 MeV) equations of state. Naturally, more is the compressibility,
more are the number of collisions and hence more positive is the directed flow.
This is indicating that the directed flow is sensitive towards the equations of state.  \\
\begin{figure}
\includegraphics{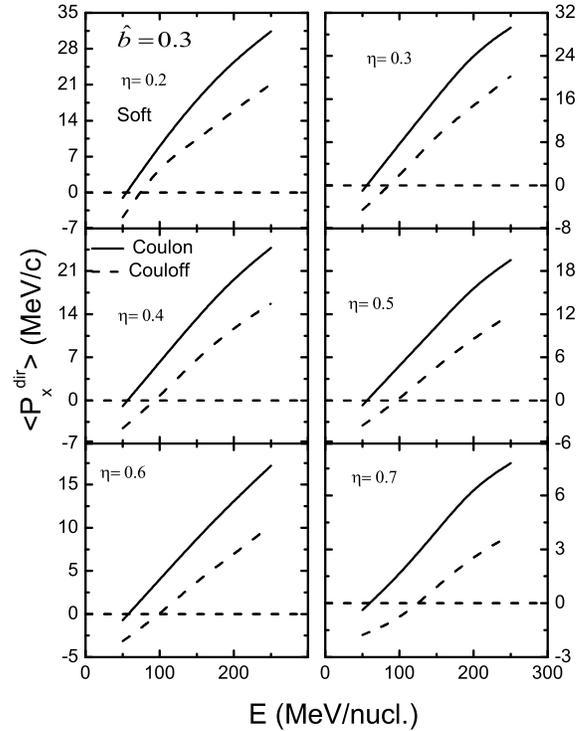}
\caption{\label{fig3} Excitation function of directed flow at different asymmetries with and without Coulomb 
interactions at semi-central geometry.}
\end{figure}
Finally, in Figs.\ref{fig3} and \ref{fig4}, we display the excitation function of directed flow at different 
asymmetries from ${\eta}$ = 0.2 to 0.7. The value of abcissa at zero value of ${<P_{x}^{dir}>}$ corresponds
to the energy of vanishing flow (EVF) or alternatively, the balance energy (${E_{bal}}$). The Fig. \ref{fig3}  
shows the shift in the balance energy due to Coulomb interactions, while, Fig. \ref{fig4} is representing the shift in the balance energy due to different equations of state. In Fig. \ref{fig3}, one sees a linear enhancement in the nuclear flow with increase in the incident energy. This increase in the transverse flow is sharp at smaller
incident energies (upto 200 MeV/nucleon). If one goes to higher incident energies, the value gets 
saturated as discussed in ref \cite{sood}. We have displayed here the results upto 250 MeV/nucleon, since we are 
interested in and around balance energy. In the presence of Coulomb interactions, more positive value of 
the flow is obtained. This is due to the well known repulsive nature of Coulomb interactions. 
At higher energies, the repulsion due to Coulomb interactions is stronger during the early phase of the
reaction and transverse momentum increases sharply. The overall effect depends on the asymmetry of the
reaction. If one looks at the balance energy, the shift in the incident energy towards the higher value is
obtained at  ${<P_{x}^{dir}> = 0}$ with the asymmetry of the reaction. This is showing that with increase
in asymmetry of the reaction and in the absence of Coulomb interactions, attractive mean-field is dominating
the large region of incident energy. The systematics of the balance energy with asymmetry of the
reaction is discussed in Fig. \ref{fig5}.\\
\begin{figure}
\includegraphics{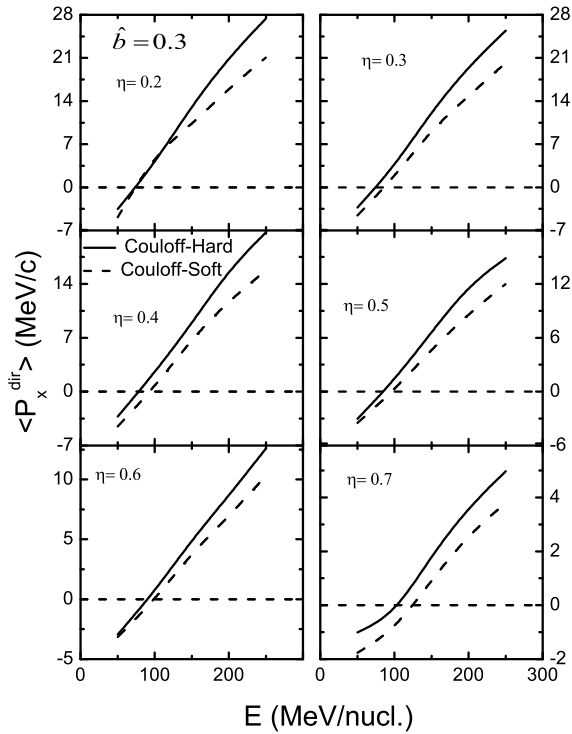}
\caption{\label{fig4} Excitation function of directed flow with hard and soft equations of state in the absence of 
Coulomb interactions for different asymmetries.}
\end{figure}
As we have seen in Fig.\ref{fig2}, that different equations of state show sensitivity towards the directed flow
with respect to the asymmetry of a reaction. The detailed analysis with soft (S) and hard (H) equations of
state is displayed in Fig. \ref{fig4}. The excitation function of directed flow follows similar
trend as explained in Fig. \ref{fig3}. For nearly symmetric systems (${\eta}$ = 0.2), the balance energy is 
found to be independent of the equations of state, however, resonable differences are observed at 
higher incident energies. More positive values of directed flow are obtained with hard equation of state
compared to soft equation of state. This is true at all asymmetries from (${\eta}$ = 0.3 to 0.7). This is due to 
different compressibilities of hard (380 MeV) and soft (200 MeV) equations of state. With an increase in
the asymmetry of a reaction, the shift in the balance energy towards the higher value of incident energy takes place with soft equation of state compared to hard one. This is consistent with the findings in the
literature \cite{rpak}. \\
\begin{figure}
\includegraphics{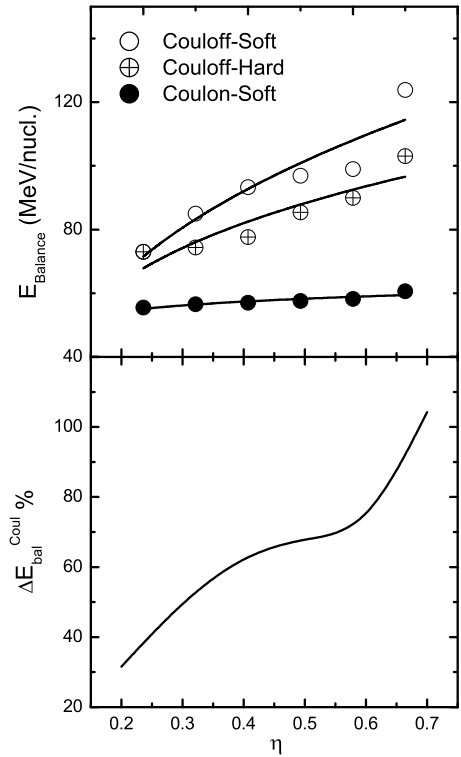}
\caption{\label{fig5} Power law dependence of balance energy with asymmetry of the reaction. The lower panel is
representing the relative ${\%}$ effect of Coulomb interactions on the balance energy.}
\end{figure}
To sum up, in Fig.\ref{fig5}, we have displayed the asymmetry dependence of balance energy. We displayed the 
results for hard and soft equations of state by switching off the Coulomb interactions. In addition 
, for a comparative study, the results in the presence of Coulomb interactions with soft equation of state are also shown. All the lines are fitted with power law of the form ${E_{bal} = C({\eta})^{\tau}}$, 
where C and ${\tau}$ are the constants. The values of ${\tau}$ in the absence of Coulomb 
interactions for soft and hard equations of state are 
0.375 and 0.282, respectively, while in the presence of Coulomb interactions for soft equation of 
state is ${\tau}$ = 0.06067.   \\
If we compare the asymmetry dependence of balance energy with mass dependence, the trend is opposite
\cite{sood}. 
It clearly indicates that if one wants to study the isospin effect, then one has to consider
the systems with ${A_{TOT} = constant}$, otherwise one is not able to get the exact information about the 
isospin effects. It is also clear from the figure, that shift in the balance energy is observed due to 
Coulomb interactions as well as due to equations of state with asymmetry of the reaction. The shift
is more due to Coulomb interactions in comparison to equations of state, indicating the importance of 
Coulomb interactions in intermediate energy heavy-ion collisions. The higher balance energy is obtained
with Coulomb-off + soft equation of state followed by Coulomb-off + hard equation of state and 
finally Coulomb-on + soft equation of state. \\
For the further understanding, the relative percentage difference in the balance energy is plotted in the
lower panel denoted by the quantity ${\Delta E_{bal}^{Coul}{\%}}$ given by\\
\begin{equation}
\Delta E_{bal}^{Coul}\% = \left[\frac{E_{bal}^{Couoff+soft} - E_{bal}^{Couon+soft}}{E_{bal}^{Couon+soft}}\right]
{\times}100
\end{equation}
The ${\Delta E_{bal}^{Coul}{\%}}$ is found to increase with the increase in the asymmetry of a reaction. 
This indicates shift of the nuclear matter towards the attractive mean-field region in the absence of 
Coulomb interactions with the asymmetry of the reaction. This difference ${\Delta E_{bal}^{Coul}}$ = 
30 MeV/nucleon at ${\eta = 0.2}$, while it is 115 MeV/nucleon at ${\eta = 0.7}$. The difference of 
90 MeV/nucleon in the shift of balance energy with asymmetry cannot be ignored. 
This is the first ever study and experiments are called to verify the results.\\        
\section {Conclusion}
Our present aim was to understand the influence of Coulomb interactions as well as equations of state
on the dynamics of large asymmetric reactions in semi-central heavy-ion collisions. 
At low incident energies, the contribution of mean-field is more for
the nearly symmetric systems, while at higher incident energies, opposite scenario
is observed. The 
balance energy is found to increase with the increase in the asymmetry of the reaction. The balance 
energy is affected by the Coulomb interactions compared to different equations of state. \\

\section {Acknowledgment}
This work has been supported by the grant from Department of Science and Technology (DST), Government of 
India, New Delhi, vide Grant No.SR/WOS-A/PS-10/2008.\\ 

\end{document}